\documentstyle[12pt]{article}
\makeatletter
\newcommand{\singlespacing}{\let\CS=\@currsize\renewcommand{\baselinestretch}
{1.0}\tiny\CS}
\newcommand{\doublespacing}{\let\CS=\@currsize\renewcommand{\baselinestretch}
{1.5}\tiny\CS}

\newcommand{\bd}{\begin{document}}
\newcommand{\ed}{\end{document}}
\newcommand{\bc}{\begin{center}}
\newcommand{\ec}{\end{center}}
\newcommand{\bfr}{\begin{flushright}}
\newcommand{\efr}{\end{flushright}}
\newcommand{\vs}{\vspace}
\newcommand{\hs}{\hspace}
\newcommand{\beq}{\begin{equation}}
\newcommand{\eeq}{\end{equation}}
\newcommand{\lb}{\linebreak}
\newcommand{\mb}{\makebox}
\newcommand{\fb}{\framebox}
\newcommand{\mc}{\multicolumn}
\newcommand{\ben}{\begin{enumerate}}
\newcommand{\een}{\end{enumerate}}
\newcommand{\bit}{\begin{itemize}}
\newcommand{\eit}{\end{itemize}}
\newcommand{\ol}{\overline}
\newcommand{\un}{\underline}
\newcommand{\lefq}{\lefteqn}
\newcommand{\ba}{\begin{array}}
\newcommand{\ea}{\end{array}}
\newcommand{\beqa}{\begin{eqnarray}}
\newcommand{\eeqa}{\end{eqnarray}}
\newcommand{\beqas}{\begin{eqnarray*}}
\newcommand{\eeqas}{\end{eqnarray*}}
\newcommand{\bfg}{\begin{figure}}
\newcommand{\efg}{\end{figure}}
\newcommand{\bds}{\begin{displaymath}}
\newcommand{\eds}{\end{displaymath}}
\newcommand{\btb}{\begin{tabbing}}
\newcommand{\etb}{\end{tabbing}}
\newcommand{\para}{\parallel}
\newcommand{\pad}{\partial}
\newcommand{\nn}{\nonumber}
\newcommand{\la}{\leftarrow}
\newcommand{\ra}{\rightarrow}
\newcommand{\lgla}{\longleftarrow}
\newcommand{\lgra}{\longrightarrow}
\newcommand{\La}{\Leftarrow}
\newcommand{\Ra}{\Rightarrow}
\newcommand{\Lra}{\Leftrightarrow}
\newcommand{\Lgla}{\Longleftarrow}
\newcommand{\Lgra}{\Longrightarrow}
\newcommand{\bm}{\boldmath}
\newcommand{\lan}{\langle}
\newcommand{\ran}{\rangle}
\renewcommand{\a}{\alpha}
\renewcommand{\b}{\beta}
\newcommand{\g}{\gamma}
\newcommand{\G}{\Gamma}
\renewcommand{\d}{\delta}
\newcommand{\eps}{\epsilon}
\newcommand{\th}{\theta}
\newcommand{\Th}{\Theta}
\newcommand{\s}{\sigma}
\newcommand{\lam}{\lambda}
\newcommand{\D}{\Delta}
\newcommand{\vare}{\varepsilon}
\newcommand{\pr}{\prime}
\newcommand{\ro}{\rho}
\newcommand{\nab}{\nabla}
\newcommand{\m}{\mu}
\newcommand{\n}{\nu}
\newcommand{\Sg}{\Sigma}
\newcommand{\p}{\pi}
\newcommand{\R}{I\!\!R}
\newcommand{\om}{\omega}
\newcommand{\Om}{\Omega}
\newcommand{\ze}{\zeta}
\newcommand{\vart}{\vartheta}
\newcommand{\tri}{\triangle}
\newcommand{\f}{\frac}
\newcommand{\iny}{\infty}
\newcommand{\pro}{\propto}
\begin{document}
\begin{center}{\large{\bf Topological Aspects of High-$T_c$ Superconductivity, 
Fractional Quantum Hall Effect and Berry Phase}}\end{center}
\begin{center}
{\bf B.Basu}\footnote{ e-mail : banasri@www.isical.ac.in}, ~ {\bf D.Pal}\
\footnote{e-mail : res9719@www.isical.ac.in}\\
{\bf and}\\ {\bf P.Bandyopadhyay}\footnote{e-mail : pratul@www.isical.ac.in}\\ 
{\bf Physics and Applied Mathematics Unit}\\
{\bf Indian Statistical Institute}\\
{\bf Calcutta-700035}
\end{center}
\date{}


\vspace*{1cm}


\centerline{\bf Abstract}

\thispagestyle{empty} 

We have analysed here the equivalence of RVB states with $\n=1/2$ FQH states
in terms of the Berry Phase which is associated with the chiral anomaly
in 3+1 dimensions. It is observed that the 3-dimensional spinons and holons
are chracterised by the non-Abelian Berry phase and these reduce to 1/2
fractional statistics when the motion is confined to the equatorial planes.
The topological mechanism of superconductivity is analogous to the topological
aspects of fractional quantum Hall effect with $\n=1/2$.

\section{Introduction}

   The revival of interest in the resonating valence bond states is due to 
the discovery of high temperature superconductors. In 1973, Anderson [1] 
suggested that compared with the classical Neel state, a better ground state
could be obtained by allowing nearby spins to couple in singlet pairs. Such
singlet bonds can exist in any lattice, in a large variety of spatial 
configurations. Fazekas and Anderson [2] substantiated this idea with s=1/2
spins on a triangular lattice. The ground state is then a sort of (quantum)
liquid of singlet bonds which can resonate among different configurations,
whence the name Resonating Valence Bond (RVB).
     
    Kivelson et al [3] analysed the structure of an RVB state on a 2D square
lattice and predicted the existence of peculiar topological elementary
excitations, having the nature of neutral, spin 1/2 excitations, roughly
described as {\it dangling bonds}, i.e. resulting from the breaking of a 
singlet pair. 
    
     In an earlier paper, Kalmeyer and Laughlin [4] have proposed that the
ground state of the frustrated Heisenberg antiferromagnet in two dimensions
and the fractional quantum Hall state for bosons might be the $same$ in the
sense that the two systems could be adiabatically evolved into one another
without crossing a phase boundary. This is compatible with the RVB concept of
Anderson to be operating in the 2D Heisenbereg antiferromagnet in a triangular
lattice where the ground state is understood to be a nondegenerate quantum
liquid  with an energy gap [1,5,6]. Kalmeyer and Laughlin [4,5] have shown that
the fractional Hall state with $\n=1/2$ confined to a triangular lattice has
the same variational energy as the Anderson RVB state. This strongly suggests
that the RVB state and FQH state with $\n=1/2$ are the same thing where the
ground state is a nondegenerate singlet and the elementary excitations are
neutral spin 1/2 fermions. In this note we want to study the topological
aspects associated with frustrated Heisenberg antiferromagnet and show the
equivalence of RVB state and FQH state with $\n=1/2$ in terms of the Berry
phase involved in the system.

     Indeed some years ago, Libby, Zou and Laughlin [7] have found that
adiabatic interchange of three dimensional spinons and holons produces nonzero
Berry phase and those reduce to 1/2 fractional statistics when motion is
confined to equatorial planes. These results discredict the importance of two
dimensionality in this type of problems and raise the possibility of monopole
superconductivity in three dimensional Mott insulators. Otherwise, noting that
most high temperature superconductors are planar materials, mechanism of
superconductivity, namely, pairing by fractional statistics [8] has thus far
been investigated only in two spatial dimensions. The discovery of quantum
disorder in any translationally invariant spin 1/2 antiferromagnet requires
one to consider the existence of neutral spin 1/2 excitations, spinon which
may be considered as carrying fractional quantum number. These in turn lead to
charged spinless excitations, called holons, which interact by means of a
gauge forces potentially capable of using superconductivity. Indeed, Libby,
Zou and Laughlin have reported the existence of such forces in three
dimensions. 

    In a series of papers [9-13], we have given a microscopic theory of FQHE in
terms of Berry phase which is associated with chiral anomaly in 3+1 dimension.
We have considered the 2DEG of N particles on the surface of a three
dimensional sphere of an infinitely large radius R in a radial (monopole)
magnetic field. As a result, the variational wave function of the homogeneous
states of incompressible fluid with finite N are constructed following
Haldane [14] and in the spirit of Laughlin [15]. The wavefunction describes a
state without gapless excitations and is equivalently characterised as an
incompressible fluid at occupations $\n=1/m$, m being an odd integer. $\n$ is
related to the Berry phase generated by the chiral symmetry breaking caused by
the external magnetic field. In the FQH state with even denominator filling
factor [11], the Berry phase is removed to the dynamical phase and we can
observe the Berry phase when the state is split into pairs of degenerate
fermions giving rise to the non-Abelian Berry phase. Indeed, it is found [12]
that a $\n=1/2$ FQH state can be represented by a pair of spin polarised
electrons in a p-wave state which has odd parity. Evidently, this resembles the
RVB state.

 In this note we shall study the topological aspects associated with frustrated
Heisenberg antiferromagnet and consider the equivalence of RVB state and FQH
  state with $\n=1/2$ in terms of the Berry phase involved in the system and 
consider high-$T_c$ superconductivity from this viewpoint. In section 2 we
shall discuss the topological aspects of Heisenberg antiferromagnet on a 
triangular lattice and RVB states. In section 3 we shall relate RVB states 
with $\n=1/2$ from the viewpoint of Berry phase and finally in section 4
 we shall consider the relationship between high- $T_c$ superconductivity
and FQH states with $\n=1/2$ in terms of non-zero Berry phase.

\section{Topological Aspects of Heisenberg Antiferromagnet on a Triangular
Lattice and Resonating Valence Bond States}
  
   We consider the antiferromagnetic Heisenberg Hamiltonian 
\beq
H=J~\sum_{i,j}~~S_i.S_j~~~~,~~~~~~~~~~~~~~~~~~~~~~~                      J>0
\eeq
where the sum is over all pairs of nearest neighbour sites of the 2D triangular
lattice and $$\vec{S_i}=\frac{1}{2}~ \hbar~ \vec{\s_i}$$ is the spin operator 
of the i-th site. The Heisenberg antiferromagnet on a triangular lattice is 
always characterised by frustration which gives rise to chirality. In our 
framework this chirality is associated with the Berry phase of our system of 
interest which is a 2D triangular lattice on the surface of a 3D sphere of 
large radius R.

    Kalmeyer and Laughlin [4] utilising the procedure of Holstein-Primakoff
transformation [16] considered the spin problem as a lattice gas by imagining
an {\it atom} to be present on every site with an up spin. The atoms 
are then bosons with creation operators 
$$a^{\dag}_j = \hbar^{-1}(S^x_j+iS^y_j)$$
The Hamiltonian (1) then becomes 
\beq
H=T+V
\eeq
where
\beq
T=\frac{1}{2}~~J~ \sum_{i,j}~~(~a^{\dag}_j a_i~ +~a^{\dag}_i a_j~)
\eeq
and
\beq
V=J~\sum_{i,j}~~(a^{\dag}_j a_i a_i a^{\dag}_j~ +~\frac{1}{2} J N_s~ -~ 6J
\sum_i~a^{\dag}_i~a_i~)
\eeq
where $N_s$ is the no. of spins or lattice sites. The boson kinetic energy
operator T comes from the spin exchange or XY part of the Heisenberg
interaction. The near neighbour repulsion of potential energy comes from the
Ising part. Now, we see that T given by eqn.(3) does not have the right free 
particle form because the hopping matrix elements $J_{ij}=J$ are positive. 
This makes the
energy bands disperse down as one moves away from the centre of the Brillouin
zone. To avoid this, they considered $J_{ij}$ to be matrix elements of right
sign, namely negetive, in the presence of a fictitious vector potential
$\vec{A}$, 
where 
\beq
\vec{A}~=~\frac{1}{2}~B~(~x\hat{y}~-~y\hat{x}~)
\eeq
with a particular value of $B$. Assigning an arbitary charge $e^*$ to the 
bosons and coupling them to $\vec{A}$, introduces phases into the matrix 
element $J_{ij}$ according to 
\beq
J_{ij}~\ra \tilde{J_{ij}}~=~-~\exp{[\frac{2{\pi}i}{\phi_0}~\int^j_i~A.ds]}
\eeq
where $\phi_0~=~\frac{hc}{e^*}$
is the quantum of flux associated with bosons of charge $e^*$.
If we choose $B$ in such a way that
\beq
\sqrt{3}~a^2_0~=~4\pi~ l^2_0
\eeq
where $a_0$ is the lattice constant and
$l_0~=~{(\frac{e^*B}{{\hbar}c})}^{-1/2}$ is the magnetic length then all
phase factors are real and corresponds to one fictitious flux quantum per
spin. For a closed loop this phase factor corresponds to the Berry phase 
when the 2D lattice gas is taken to reside on the surface of a 3D sphere in 
a radial (monopole) magnetic field.

    It may be observed here that the Berry phase is associated with the chiral
anomaly which is caused by quantum mechanical symmetry breaking when a chiral
current interacts with a gauge field. Indeed, the divergence of the axial
vector current in the quantum mechanical case does not vanish and is
associated with the topological quantity known as the Pontryagin index through
the relation 
\beq
q~=~-\frac{1}{16 {\pi}^2}~\int~Tr ^*F_{\m\n}~F_{\m\n}~d^4x
~=~-\frac{1}{2}~\int \partial_\m~J^5_\m~d^4x
\eeq
where $J^5_\m$ is the axial vector current and $q$ is the Pontryagin index. The
Pontryagin index associated with the integral of the chiral anomaly is related
to the Berry phase which arises when a quantum particle described by a
parameter dependent Hamiltonian moves in a closed path. Indeed, the Berry
phase acquired by such a particle is given by $e^{i{\phi}_B}$ where 
\beq
{\phi}_B~=~2\pi \m~=~\pi q
\eeq
with the relation $q~=~2\m$ [17]. Here $\m$ appears as a monopole strength.

    In a 3D anisotropic space, we can construct the spherical harmonics
$Y^{m,\m}_l$ with $l~=~1/2$, $|m|=|\m|=1/2$ when the angular momentum 
relation is given by 
\beq
\vec{J}~=~\vec{r}~\times \vec{p}~-~\m\vec{r}
\eeq
where $\m$ can take the values as $0,\pm 1/2,\pm 1,\pm 3/2, \ldots$
This is similar to the angular momentum relation in the case when a charged
particle moves in the field of a magnetic monopole. Fierz [18] and Hurst [19]
have studied the spherical harmonics $Y^{m,\m}_l$. Following them we can write
\beq
Y^{m,\m}_l~=~(1+x)^{-\frac{(m-\m)}{2}}(1-x)^{\frac{-(m+\m)}{2}}.
\frac{d^{l-m}}{d^{l-m}x}~[(1+x)^{l-\m}(1-x)^{l+\m}]~e^{im\phi}~e^{-i\m\chi}
\eeq
where $x~=~cos{\th}$ and the quantity m and $\m$ just represent the
eigenvalues of the operator $i\frac{\partial}{\partial{\phi}}$ and
$i\frac{\partial}{\partial{\chi}}$ respectively. It is noted that apart from
the usual angles $\th$ and $\phi$, we have an extra angle $\chi$ denoting the
rotational orientation around a specified fixed axis attached to a space-time 
point $x_\m$ giving rise to an anisotropy
in the space-time manifold [20]. The fact that in such an anisotropic space
the angular momentum can take the value 1/2 is found to be analogous to the
result that a monopole charged particle composite representing a dyon
satisfying the condition $e{\m}~=~1/2$ has its angular momentum shifted by
1/2 unit and its statistics shifted accordingly [21].

    Now when a 2D triangular lattice is taken to reside on the surface of a 3D
sphere of large radius in a radial (monopole) magnetic field, we can associate
the chirality with the Berry phase and the antiferromagnetic Heisenberg
Hamiltonian can be represented by an anisotropic Hamiltonian [22].
\beq
H~=~J~\sum~(~S^x_iS^x_{i+1}~+~S^y_iS^y_{i+1}~+~\Delta ~S^z_iS^z_{i+1})
\eeq
where $J~>~0$ and $\Delta ~\geq~0$, $\Delta$ the anisotropic parameter is given
by $\Delta~=~\frac{2\m+1}{2}$ where $\m$ is related to the Berry phase
factor. It is noted that $\Delta~=~1$ one has the antiferromagnetic
Heisenberg model and for $\Delta~\ra~\infty$ one has the Ising model. When
$\Delta~=~0$, we have the XX model. It may be mentioned here that the
relationship of the anisotropic parameter $\Delta$ with the Berry phase factor
$\m$ has been formulated from an analysis of the relationship between the
conformal field theory in 1+1 dimension, Chern-Simon theory in 2+1 dimension
and chiral anomaly in 3+1 dimension [22]. Indeed the association of conformal
field theory with quantum group having the deformation parameter $q$ of the 
deformed algebra $U_q(SL(2))$ helps us to relate the deformation parameter $q$
with the Berry phase factor $\m$. In the framework of quantum group, we can 
consider the following open chain Hamiltonian [23]
\beq
H_N(q,y)=\f{J}{4}\left( {\sum_{i=1}}^{N-1}\sigma_i^x~\sigma^x_{i+1}
+\sigma_i^y~\sigma^y_{i+1}+\f{q+q^{-1}}{2}\sigma^z_i~\sigma^z_{i+1}
-\f{q-q^{-1}}{2}(\sigma_1^z-\sigma^z_N)\right)
\eeq
which maintains quantum group symmetry $SU_q(2)$. 
In the limit $q\to 1$, one recovers the usual $SU(2)$ algebra. Now if we set
$\Delta=\f{q+q^{-1}}{2}$, we see from eqns.(12) and (13) that the bulk terms
of (13) and the one in (12) coincide. The only difference appears in the 
boundary term of (13) which is essential for the quantum group symmetry.
As $\m=1/2$ represents a free fermion case corresponding to $q=1$, we note that
in the Hamiltonian (13) (neglecting the boundary term) if we choose
$\f{q+q^{-1}}{2}~=~\f{2\m+1}{2}$, we get the isotropic Heisenberg model.

    Now for a 2D triangular lattice on the surface of a 3D sphere in a radial
magnetic field, the chirality demands that $\m$ is non-zero and should be
given by $|\m~|=~1/2$ in the ground state. However, for $\m~=~1/2$, we
get the antiferromagnetic Heisenberg Hamiltonian without frustration. So a
frustrated spin system should be given by $\m~=~-1/2$ suggesting
$\Delta~=~\frac{2\m+1}{2}~=~0$ in the Hamiltonian (12). But with
$\Delta~=~0$, this Hamiltonian effectively represents the XX model
corresponding to a bosonic system represented by singlets of spin pairs. This
eventually leads to the resonating valence bond state giving rise to a
nondegenerate quantum liquid.

\section{Resonating Valence Bond, Fractional Quantum Hall Effect with
$\n=1/2$ and Berry Phase} 

    We have shown that when an anisotropic Heisenberg Hamiltonian of an
antiferromagnetic spin system is characterised by a particular value of Berry
phase factor related to a fixed value of anisotropic parameter, then the 
system leads to the
generation of quantum spin liquid corresponding to a resonating valence bond
state. The RVB state is characterised by the anisotropic parameter 
 related to the particular value of Berry phase. We point out here that the 
spin singlet states forming the quantum
liquid are equivalent to FQH liquid  with filling factor $\n~=~1/2$. Indeed,
 in some earlier papers [9-13], we have pointed out that in QHE the external
magnetic field causes a chiral symmetry breaking of the fermions (Hall
particles) and as a result an anomaly is realised in association with the
quantization of Hall conductivity. This helped us to study the behaviour of a
quantum Hall fluid from the viewpoint of the Berry phase which is linked with
chiral anomaly. We consider a 2DEG of N-particles on the spherical surface of
a 3D sphere in a radial (monopole) strong magnetic field. From the description
of spherical harmonics $Y^{m,\m}_l$ given before we can construct spinor
$$\th~=~\left(\begin{array}{c}u\\v\end{array}\right)$$ where
$$u~~=~~Y^{1/2,1/2}_{1/2}~=~sin{(\th /2)}~\exp{\frac{i(\phi - \chi)}{2}}$$
\beq
v~=~Y^{-1/2,1/2}_{1/2}~=~cos{(\th /2)}~\exp{\frac{-i(\phi + \chi)}{2}}
\eeq
The N-particle wave function is then written as :
\beq
\psi^{(m)}_N~=~\prod_{i<j}~{(u_iv_j~-u_jv_i)}^m
\eeq
where m is the inverse of the filling factor $\n$. Evidently, for odd (even)
m, we will have the fermionic (bosonic) states. Following Haldane [14] we
identified $m~=~J_{ij}~=~J_i~+~J_j$ for an N-particle system. It is noted
that the $m~=~1$ state which describes the complete filling of the lowest
Landau level corresponds to the ground state for the contribution of the
factor $\vec{r}~\times~\vec{p}~=~0$ with $\m~=~1/2$. Indeed, following the
Dirac quantization condition $e{\m}~=~1/2$ we note that the quasiparticle
for $m~=~(\frac{1}{\n})~=~1$ will exhibit the IQHE with fermion number 1.
However, if we consider the state with $\vec{r}~\times~\vec{p}~=~1$, the
respective angular momentum is changed to $J~=~3/2 $ for $\m~=~1/2$. This
can be viewed as a system with $\m_{eff}~=~3/2$ having
$\vec{r}~\times~\vec{p}~=~0$. This will give rise to
$m~=~\frac{1}{\n}~=~3$ when the fermion number of the quasiparticle is found
to be 1/3 which is evident from the relation $e{\m_{eff}}~=~1/2$. In this
way, we can study all FQH states with $\n~=~1/m$, m being an odd integer
given by $m~=~2\m_{eff}$. For states with $\n~=~\frac{n}{2\m_{eff}}$, we
consider the higher Landau level with the Dirac quantization condition
$e\m~=~\frac{1}{2}n$ where n can be viewed as a vortex of strength 2l+1.
The states with $\n~=~\frac{n^{\pr}}{2\m_{eff}}$, $n^{\pr}$ being an even 
integer, can be generated through conjugate states [11,13].

    This analysis suggests that for the FQH fluid with even denominator
filling factor we face a peculiar situation. For example, for the FQH state 
with $\n~=~1/2$ the Dirac quantization condition $e~\m~=~1/2$ suggests that
$\m~=~1$. Then in the angular momentum relation
$$\vec{J}~=~\vec{r}~\times~\vec{p}~-~\m\vec{r}$$ for $\m~=~1$ (or an
integer) we can use a transformation which effectively suggests that we can
have a dynamical relation of the form 
\beq
\vec{J}~=~\vec{r}~\times~\vec{p}~-~\m \vec{r}~~=~~\vec{r^\pr}~\times~
\vec{p^\pr}
\eeq
This equation indicates that the Berry phase which is associated with $\m$ may
be unitarily removed to the dynamical phase. This implies that the average
magnetic field may be taken to be vanishing in these states. However, to
observe the effect of the Berry phase, we can split the state into a pair of
electrons, each with the constraint of representing the state $\m~=~\pm
1/2$. Now as $|\m|~=~1$ is achieved for a pair of electrons each having
either $\m~=~+1/2$ or $-1/2$. We can construct the two-component spinor 
$$\th~=~\left(\begin{array}{c}u\\v\end{array}\right)$$
 from $Y^{m,\m}_l$ with $|\m|~=~|m|~=~1/2$ where
\beq
u~=~Y^{1/2,1/2}_{1/2}~~~~~,~~~~~v~=~Y^{-1/2,1/2}_{1/2}
\eeq
The charge conjugate states are 
\beq
\tilde{u}~=~Y^{-1/2,1/2}_{1/2}~~~~~,~~~~~\tilde{v}~=~Y^{1/2,-1/2}_{1/2}
\eeq
    This indicates that electrons in the pair have a definite chirality given
by $\m=1/2(-1/2)$. So each electron in the pair is spin polarised. Thus 
for $|\m|~=~1$ we can depict the pair as a p-wave state of spin
polarised electrons [24]. 
    
    From this analysis, it appears that these pairs will give rise to the
SU(2) symmetry, as we can consider the state of these two electrons as a SU(2)
doublet. We know that the usual Laughlin state for a quantum Hall fluid can be
represented by the Abelian U(1) symmetry having an Abelian Berry phase. Due to
pairing, in the even denominator quantum Hall states we will have
$SU(2)~\times~U(1)$ symmetry indicating that there will be a non-Abelian Berry
phase. Thus these states will represent non-Abelian Hall fluid. 
   
    It may be observed that for an odd parity p-wave paired FQH state for spin
singlets of polarised electrons, the total wave function is given by the
product of the Pfaffian wave function $\phi_{pf}$ and the Laughlin wave
function 
$\phi_m$ as \beq
\phi_p~=~\phi_{pf} \phi_m
\eeq
\beq 
\phi_{pf}~=~{\cal A}~\left(\frac{1}{{(z_1-z_2)}~{(z_3-z_4)}~........}\right)
\eeq
\beq
\phi_m~=~\left( \prod_{i<j}{(z_i~-~z_j)}^m~exp{({-1/4} \sum {|z_i|}^2)}
\right) 
\eeq
where $z_i~=~x_i~+~iy_i$ and $\cal A$ is the antisymmetrization operator and
$m~=~\frac{1}{\n}~=~2$. It has been shown that the non-Abelian part of
the bulk wave function $\phi_{pf}$ can be written as a correlation of the
primary fields $\psi$ in the Ising model where $\psi$ represents the free
fermion field in a free Majorana fermion theory. In the Berry phase
formalism each fermion is characterised by $|\m|~=~1/2$ forming the
pair. The Abelian part $\phi_m$ with $m~=~2$ represents the singlet formed
by a spin pair and corresponds to a bosonic case. 
    
    These odd parity singlets may be taken to be equivalent to spin singlets
of RVB states generating a quantum liquid. Indeed, we may observe that a many
body system comprising the pairs of spin singlet states will give rise to an
antiferromagnetic chain. In other words, an anisotropic antiferromagnetic
Heisenberg model with a particular value of anisotropic parameter
$\Delta~=~0$, (which corresponds to a particular value of Berry phase
factor) will effectively give rise to the RVB states generating the quantum
spin liquid. Equivalently the anisotropic Heisenberg Hamiltonian (12) which
reduces the bosonic XX model with anisotropic parameter $\Delta~=~0$, 
represents the FQH bosonic state with $\n~=~\frac{1}{m}~=~1/2$. The
correlation of FQH states with $\n~=~1/2$ and RVB spin singlet states
 can be represented in our formalism as 
\beq
\psi_{\n=1/2}~=~\phi_{pf}~\psi^{m=2}_{N}~\equiv~\psi_{RVB}
\eeq
where $\phi_{pf}$ and $\psi^{m=2}_{N}$ is given by (20) and (15) respectively. 

\section{High $T_c$ Superconductivity, Fractional Quantum Hall Effect with 
$\n~=~1/2$ and Berry phase}

    The equivalence of RVB states with $\n~=~1/2$ FQH states are
characterised by neutral spin 1/2 excitations called spinons and charged
spinless excitations called holons. To study these excitations in the
framework of our present analysis, we note that a single spin down
electron at a site j is surrounded by an otherwise featureless spin liquid
representing a RVB state. The 3D formulation of such a system will be taken to
be such that 2D frustrated spin system lies on the surface of a 3D sphere in 
a radial (monopole) magnetic field which gives rise to the
chirality associated with the magnetic monopole. This in turn gives rise to
the Berry phase factor $\m~=~-1/2$. Now the single spin down state
characterised by $\m~=~-1/2$ when coupled with this $\it {monopole}$
represented by $\m~=~-1/2$ will give rise to a state having $\m~=~-1$. As
described in previous section, the Berry phase factor $|\m|~=~1$
effectively gives rise to a FQH state with $\n~=~1/2$. Indeed, the state
characterised by $|\m|~=~1$ is formed by the single spin state
$(\m~=~1/2)$ in the spin liquid and the $\it{orbital~spin}$ caused by the
$\it{monopole}$ represented by the $\m~=~-1/2$ characteristic of a triangular
lattice in three dimensions. In this framework, the neutral spin 1/2
excitation, the spinon is such that, the elementary spin 1 excitation
characterised by $|\m|~=~1$ may split into two parts, with one spin 1/2
excitation in the bulk and the other part is due to the $\it{orbital~spin}$ 
which is in the background characterised by the chirality of a triangular 
lattice. This is analogous to the idea of Laughlin [5,6] that spinons obey 
spin 1/2 statistics. 

    Now when a hole is introduced into the system by doping, this may combine
with this spinon giving rise to a spinless charged excitation known as holons.
In these frame work holons may also be represented by FQH liquid $\n~=~1/2$
corresponding to a singlet characterised by a flux $\phi_0~=~\frac{hc}{2e}$.
Obviously these resemble Cooper pairs. As these states correspond to FQH
liquid with $\n~=~1/2$, the ground state will represent a quantum liquid
with an energy gap. This corroborates with the idea of Laughlin [25] that a 
gas of such particles might actually be a superconductor with a charge 2 order
parameter. As stated by Laughlin [25], the analog of the fractionally charged
quasiparticle is the {\it spinon}, the analog of the compressional sound wave
is an {\it antiferromagnetic spin wave}, the analog of Wigner crystalisation
is {\it antiferromagnetic ordering} and the analog of the magnetoroton gap is
a {\it magnetic fluctuation gap}. The gap of the spin wave spectrum of the
magnet gives the measure of how {\it liquid} the spin liquid is.

    It may added that Weigmann [26,27] has studied topological mechanism of
superconductivity where it is argued that strongly correlated electronic
systems represent physical systems where topological fluid may appear. Noting
that there are two operators which characterize the ground state of an
antiferromagnet, namely density of energy
\beq
\epsilon_{ij}~=~(1/4~+~\vec{S_i}.\vec{S_j})
\eeq
and chirality or measure of topological order
\beq
W(C)~=~Tr~\prod_{i{\in} C}(1/2~+~\vec{\s}.\vec{S_i})
\eeq
where $\s$ are Pauli matrices and C is the lattice contour. Weigmann has
related these operators with the amplitude and phase $\Delta_{ij}$ of
Anderson's RVB through
\beq
\epsilon_{ij}~=~{|\Delta_{ij}|}^2
\eeq
\beq
W(C)~=~\prod_C~\Delta_{ij}
\eeq
This suggests that $\Delta_{ij}$ is a gauge field. It can be locally
transformed by a U(1) transformation 
\beq
\Delta_{ij}~\rightarrow~\Delta_{ij}~e^{i(\a_i~-~\a_j)}
\eeq
The topological order parameter W(C) aquires the form of a lattice Wilson loop
\beq
W(C)~=~e^{i\phi (C)}
\eeq
 which is associated with the flux of the RVB field 
\beq
e^{i\phi(C)}~=~\prod_C~e^{iA_{ij}}
\eeq
where $A_{ij}$ is a phase of $\Delta_{ij}$ representing a magnetic flux which
penetrates through a surface enclosed by the contour C. 
 This phase is essentially the Berry phase related to chiral anomaly when 
we describe the system in three dimensions through the relation
\beq
W(C)=~e^{i 2 \pi \m}
\eeq
when the 2D surface is considered to be the surface of a 3D sphere of large
radius in a radial magnetic field.
 Infact, we have formulated here topological
superconductivity and showed that it is associated with the topological
aspects of FQH fluid with $\n~=~1/2$.

\section{Discussion}
Kalmeyer and Laughlin [4] have pointed out the equivalence of the ground
states of the frustrated Heisenberg antiferromagnet in 2D and FQH states with
$\n~=~1/2$ in the sense that the two system could be adiabatically evolved
into one another without crossing a phase boundary. We have studied here the
topological aspects of RVB states in terms of the Berry phase and have shown
its relationship with the topological aspects of FQH states with $\n~=~1/2$.
Indeed, we have shown here that three dimensional spinons and holons produce
nonzero Berry phase and these reduce to 1/2 fractional statistics when motion
is confined to equatorial phases. This is in confirmity with the idea of
Laughlin. In view of this we have interpreted here high $T_c$ superconductivity
as topological superconductivity and is not just confined to 2D planes. In
fact, there is a possibility that 3D Mott insulators may also produce
superconductivity with proper doping. Weigmann [27] have argued that 
topological superconductivity - a result of developing of topological order 
- takes place in
some models of highly correlated electron systems in three dimensions. We shall
study the Meissner effect in this framework in our further report.

It may be pointed out here that the equivalence of RVB states with FQH states
with $\n~=~1/2$ also raises the possibilities that high $T_c$
superconductivity may be studied in the framework of conformal field theory as
the FQH states are being studied from this viewpoint also. In fact we have
noted that RVB states are formed when the anisotropic parameter $\Delta
~=~\frac{2\m~+~1}{2}$ in the Hamiltonian (12) is zero indicating that
$\m~=~-~1/2$. Now from an analysis of the relationship between the conformal
field theory in 1+1 dimension, Chern-Simons theory in 2+1 dimension and
chiral anomaly in 3+1 dimension a relationship between the central charge c
in conformal field theory and the Berry phase factor $\m$ associated with the
chiral anomaly may be established and it is found to be given by $c~=~1-\frac
{6}{m(m+1)}$ with $m~=~2\m ~+~2$ [22]. This suggests that $\m~=~1/2$
effectively corresponds to $c~=~-2$. Indeed, the non-Abelian part of the wave
function for some FQH states with even denominator filling factor have been
studied in the framework of conformal field theory with $c~=~-2$.  It may be
added here that there exists a class of conformal field theories with a chiral
algebra which may be associated with Wess-Zemino-Witten (WZW) theories. In
view of this, we may have another version of topological superconductivity in
terms of WZW theories [28].

{\bf Acknowledgement} : One of the authors (P.B.) is grateful to the Council
of Scientific and Industrial Research (CSIR) for a financial grant (scheme no.
21{0377)/96/EMR-II) supporting this research. In addition D.P. is also
grateful to CSIR for their financial support in doing this work.

\newpage
\begin{center}
{\bf References}
\end{center}
\begin{enumerate}
\item P.W. Anderson, Mat. Res. Bull., {\bf 8} (1973), 153.
\item P. Fazekas and P.W. Anderson, Phil. Mag., {\bf 30} (1974), 432.
\item S.A. Kivelson, D.S. Rokhsar and J.P. Sethna, Phys. Rev. {\bf B, 35}
(1975),  8865.
\item V. Kalmeyer and R.B. Laughlin, Phys. Rev. Lett, ${\bf 59}$ (1987), 2095.
\item R.B. Laughlin, Phys. Rev. Lett, ${\bf 50}$ (1983), 1395
\item R.B. Laughlin, in {\it Quantum Hall Effect}, Eds. S.M. Girvin and R.E.
 Prange 
\item[] (Springer-Verlag, NewYork (1986), p. 233).
\item S.B. Libby, Z. Zou and R.B. Laughlin, Nucl. Phys. ${\bf B, 348}$ (1991),
 693.
\item Y. Chen, F. Wilczeck, E. Witten and B.I. Halperin, Int. J. Mod. Phys.
{\bf B, 3} (1989), 1001.
\item D. Banerjee and P. Bandyopadhyay, Mod. Phys. Lett. ${\bf B, 8}$ (1994),
 1643.
\item B. Basu, Mod. Phys. Lett. ${\bf B, 6}$ (1992), 1601.
\item B. Basu and P. Bandyopadhyay, Int. J. Mod. Phys. ${\bf B, 11}$ (1997), 
2707.
\item B. Basu, D. Banerjee and P. Bandyopadhyay, Phys. Lett. ${\bf A, 236}$
 (1997), 125.
\item B. Basu and P. Bandyopadhyay, Int. J. Mod. Phys. ${\bf B, 12}$ (1998), 
49.
\item F.D.M. Haldane, Phys. Rev. Lett. ${\bf 51}$ (1983), 605.
\item R.B. Laughlin, Surf. Sci. ${\bf 142}$ (1984), 163.
\item T. Holstein and H. Primakoff, Phys. Rev. ${\bf 58}$ (1940), 1098.
\item D. Banerjee and P. Bandyopadhyay, J. Math. Phys. ${\bf 33}$ (1992), 990.
\item M. Fierz, Helv. Phys. Acta. ${\bf 17}$ (1944), 27.
\item C.A. Hurst, Ann. Phys. ${\bf 50}$ (1968), 51.
\item P. Bandyopadhyay, Int. J. Mod. Phys. ${\bf A4}$ (1989), 4449.
\item F. Wilczeck, Phys. Rev. Lett. ${\bf 48}$ (1982), 1146.
\item P. Bandyopadhyay, Conformal Field Theory, Quantum Groups and Berry Phase
(submitted for publication).
\item J. Gonzalez, M. Martin-Delgado, G. Sierra, A.H. Vozmediano, {\it Quantum 
Electron Liquids} {\it and High-$T_c$ Superconductivity} (Springer, 1995).
\item G. Moore and N. Read, Nucl. Phys. ${\bf B, 360}$ (1991), 362.
\item R.B. Laughlin, Science {\bf 242} (1988), 525.
\item P. Wiegmann, in {\it Field Theory}, {\it Topology and} {\it Condensed
Matter Physics},  
\item[] Ed. H.B. Geyer (Springer, 1995) p.177.
\item P. Wiegmann, Prog. Theor. Phys. Suppl. ${\bf 107}$ (1992), 243. 
\item X.G. Wen and Y.S. Wu, Nucl. Phys. ${\bf B, 419}$ (1994), 455.

\end{enumerate}
\ed